\documentclass[twocolumn, superscriptaddress, secnumarabic, amssymb, showpacs, nobibnotes, aps, prb]{revtex4-2}
\usepackage{graphicx }
\usepackage{dcolumn}
\usepackage{bm}
\usepackage{xcolor}
\usepackage{upgreek}
\usepackage{xspace}
\usepackage{amsmath}
\usepackage[]{hyperref}
\hypersetup{colorlinks=true,linkcolor=blue,citecolor=blue,urlcolor=blue,pdfpagemode=UseNone}

\begin{document}
\newcommand{\rubr}{RuBr$_3$}
\newcommand{\muSR}{$\mu$SR\xspace}
\newcommand{\mus}{$\upmu$s\textsuperscript{-1}\xspace}
\newcommand{\REF}{{\bf[REF]}}

\title{Magnetism in Kitaev Quantum Spin Liquid Candidate RuBr$_3$}
\author{T. Weinhold}
\affiliation{Institute of Solid State and Materials Physics, TU Dresden, D-01062 Dresden, Germany}
\author{C. Wang}
\affiliation{Laboratory for Muon Spin Spectroscopy, PSI, Villigen, Switzerland}
\author{F. Seewald}
\affiliation{Institute of Solid State and Materials Physics, TU Dresden, D-01062 Dresden, Germany}
\author{V. Grinenko}
\affiliation{Tsung-Dao Lee Institute, Shanghai Jiao Tong University, Shanghai, China}
\author{Y. Imai}
\affiliation{Department of Physics, Graduate School of Science, Tohoku University, 6-3, Aramaki Aza-Aoba, Aoba-ku, Sendai, Miyagi 980-8578, Japan}
\author{F. Sato}
\author{K. Ohgushi}
\affiliation{Department of Physics, Graduate School of Science, Tohoku University, 6-3, Aramaki Aza-Aoba, Aoba-ku, Sendai, Miyagi 980-8578, Japan}
\author{H.-H. Klauss}
\author{R. Sarkar}
\email{rajib.sarkar@tu-dresden.de}
\affiliation{Institute of Solid State and Materials Physics, TU Dresden, D-01062 Dresden, Germany}

\date{\today}

\begin{abstract}
The present studies show that long-range magnetic order takes place in RuBr$_3$ at $\approx$ 34\, K. The observations of clear oscillations in the muon time spectra demonstrate the presence of well-defined internal fields at the muon sites. The magnetic ordering appears to be very robust and static suggesting a more conventional nature of magnetic ordering in the RuBr$_3$ system at zero field. Present investigations prove that in RuBr$_3$ the Kitaev interactions are likely to be weakened at zero field in comparison to the $\alpha$-RuCl$_3$ system. This proves that it is possible to tune the Kitaev interactions by replacing Cl with heavier halogen elements such as Br.

\end{abstract}

                          
\maketitle
\textit{Introduction.} 
A quantum spin liquid (QSL) is an exotic state of matter, in which electron spins are strongly entangled, but do not exhibit any long-range magnetic ordering down to $T$\,=\,0\,K. Despite considerable effort in the past, so far, experimental realizations of a clean QSL remain scarce.~\cite{{Balents2010},{Balents-2017},{Wen2019}} 
Alexei Kitaev proposed a theoretical model, known as the Kitaev model, where spins (S\,=\,1/2) are placed on a honeycomb lattice. They are coupled with their three nearest neighbor spins with bond-dependent ferromagnetic Ising interactions.~\cite{KITAEV20062} Kitaev model is exactly solvable, and the ground state is a quantum spin liquid. Recently the ruthenium halides are considered promising candidates for the realization of a Kitaev spin liquid.  The bond-dependent Kitaev magnetic exchange interactions drive this quantum spin liquid phase. To realize this experimentally,  $\alpha$-RuCl$_3$ has been extensively studied. However, the theoretical descriptions of $\alpha$-RuCl$_3$ are still debated, and a correct microscopic description of the low-temperature phase is not available. \cite{{PhysRevLett102017205},{Winter2017},{Motome2020},{PhysRevB92235119}, {PhysRevB93134423},{Banerjee2016fb},{PhysRevResearch2033011},{PhysRevLett119037201},{Kasahara2018dl}} It appears that a lack of analog materials for $\alpha$-RuCl$_3$ puts forward certain limitations to find out the electronic properties as functions of interaction parameters. In $\alpha$-RuCl$_3$ the covalency between Ru-4d and Cl-p-orbitals is important to induce Kitaev interaction. Thus, replacing Cl with heavier halogen elements, such as Br or I, is a sensible approach to strengthen the Kitaev interaction further. Indeed, a recent comparative study of the electronic structures of $\alpha$-RuCl$_3$,  RuBr$_3$, and RuI$_3$ and the estimation of magnetic exchange interactions of all three compounds suggests that RuBr$_3$ is a promising candidate to realize Kitaev spin liquid phases.~\cite{{PhysRevB.105.L041112},{KimASCT2021},{NAWAJPSJ2021},{Kaib2022}}
\\
The material RuBr$_3$ has a BiI$_3$-type structure (space group R$\bar{3}$) where Ru$^{3+}$ forms an ideal honeycomb lattice even at room temperature and it does not show a structural transition down to low temperatures. RuBr$_3$ orders antiferromagnetically at $\approx$ 34\,K. 

To further investigate this system, in particular, its nature of the static and/or dynamic ground state, we have performed detailed Muon Spin Resonance (\muSR) experiments, both in zero-field (ZF) and in the longitudinal field (LF) along the initial muon polarisation, in the temperature range 1.5-200\,K. Present $\mu$SR studies confirm the presence of long-range magnetic ordering at $\approx$\,34\,K. Muon relaxation rates ($\lambda$) probe the dynamic/static spin susceptibility at $\mu$eV energy scales. This gives important information which is complementary to nuclear magnetic resonance and inelastic neutron scattering. Present data are typical representations of static magnetism.  Thus in zero-field, the magnetic ordering is robust and static suggesting a more conventional type of quasi three-dimensional magnetic ordering in RuBr$_3$ at least in the \muSR time scale.  
However, the field dependence of the interactions is not yet known.

\begin{figure}[h]
\includegraphics[width=\columnwidth]{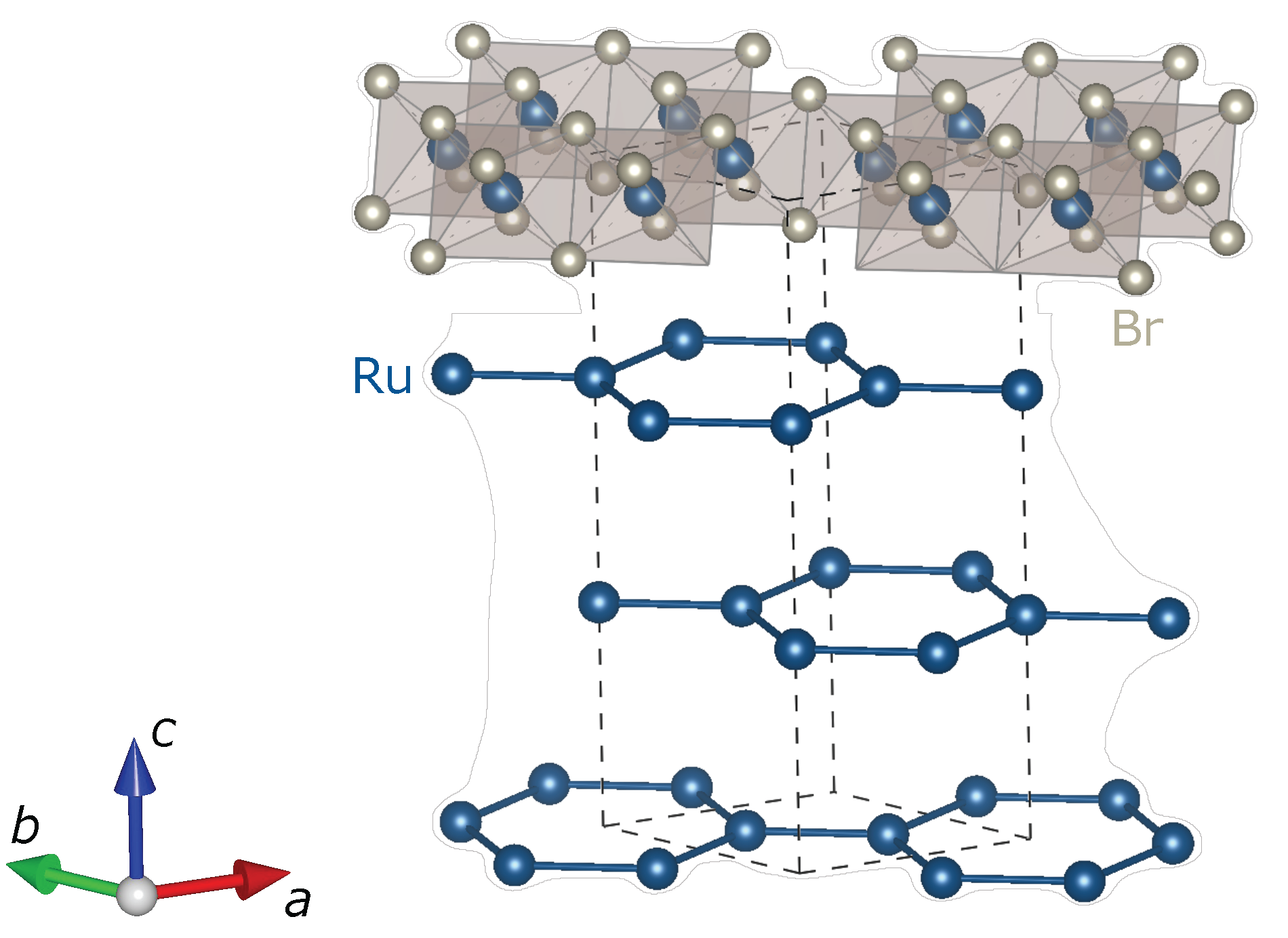}
\caption{\label{fig:ZF-musr-spectra} Crystal structure of the RuBr$_3$ system.}
\end{figure}

\begin{figure}[h]
\includegraphics[width=.5\columnwidth]{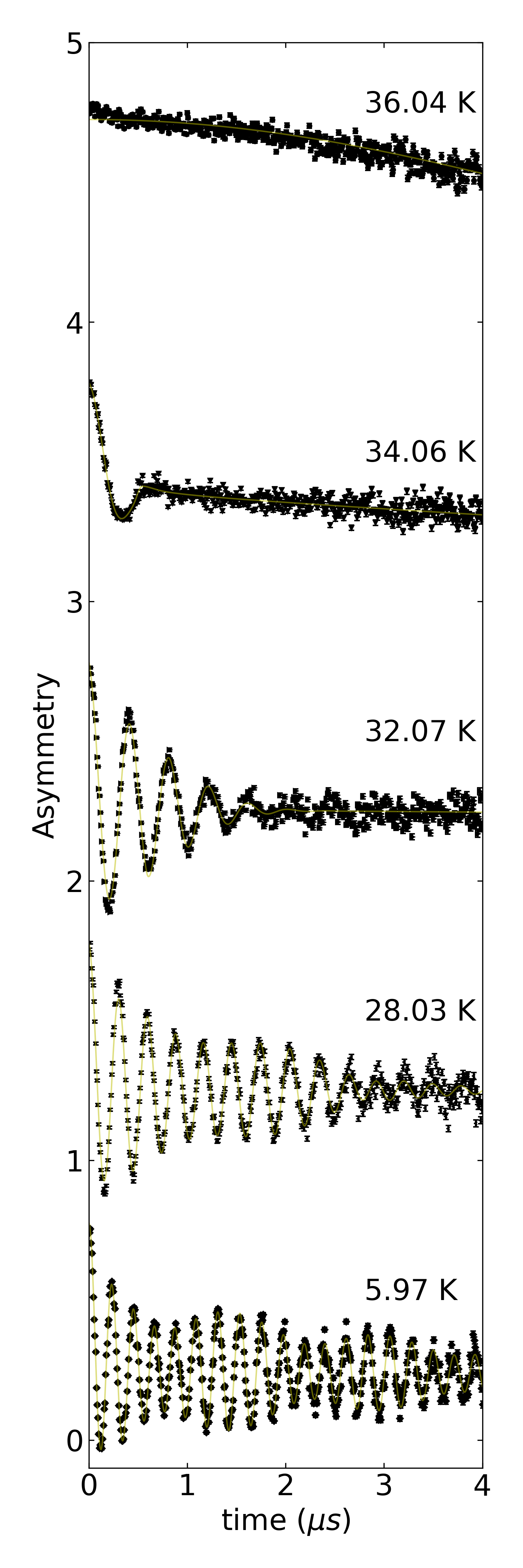}\hfill
\includegraphics[width=.5\columnwidth]{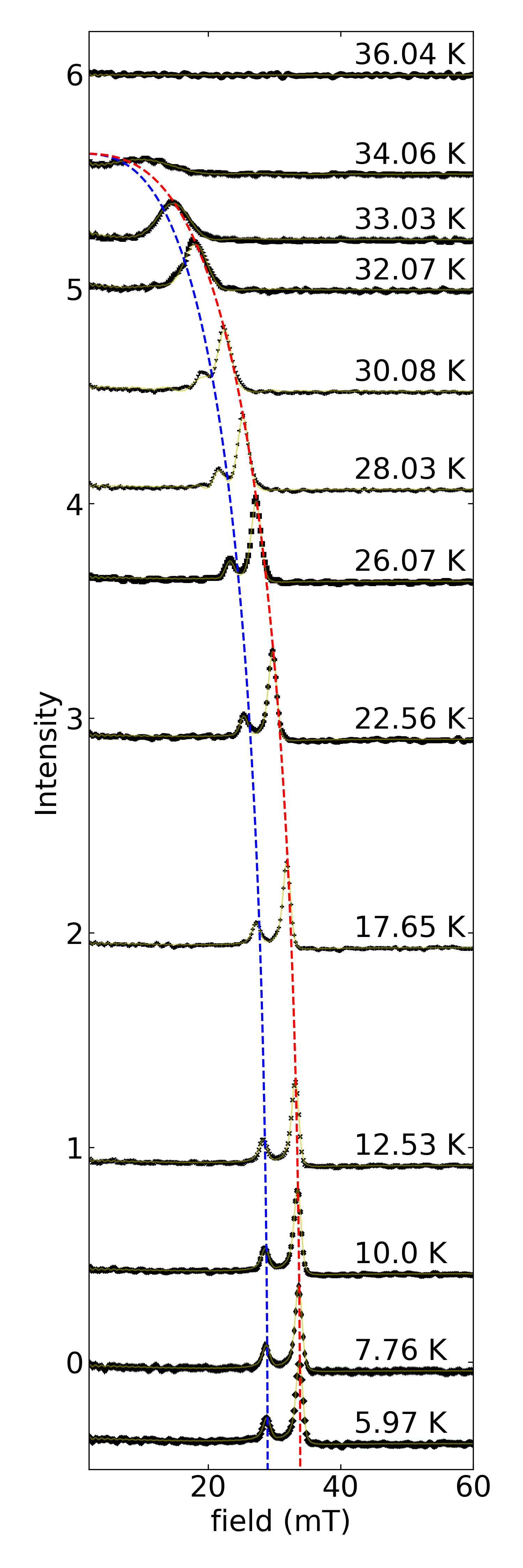}
\caption{\label{fig:ZF-musr-spectra} Left panel: At representative temperatures true ZF-\muSR\ time spectra measured at PSI . Right panel: Fourier transform \muSR\ time spectra at zero fields at a selected temperature. Fourier-transformed spectra are vertically displaced for clear demonstrations.   
Lines indicate the theoretical description as detailed in the text.}
\end{figure}

\textit{Experimental.} Polycrystalline sample of RuBr$_3$  was prepared by a high-pressure synthesis according to Ref~[\onlinecite{PhysRevB.105.L041112}]. \muSR\ experiments were performed at the 
PSI, SWISS using the GPS instrument. For muon measurements, 300\,mg of a powder sample was used. The \muSR\ data were analyzed with the muSR fit program.~\cite{sutermuon} The crystal structure of the RuBr$_3$ system is drawn using VESTA.~\cite{Momma:db5098}

\textit{\muSR\ results: the presence of long-range magnetic ordering.} Representative ZF-$\mu$SR asymmetry spectra, measured in wide temperature ranges, are shown in the left panel of Fig.~\ref{fig:ZF-musr-spectra}. The right panel shows the Fourier transformed $\mu$SR time spectra at zero fields at the selected temperature. For visual clarity, both time spectra and the Fourier-transformed spectra are shifted vertically. In general, implanted muons (here positive muons) are highly sensitive to the local magnetic fields with a resolution about ($B = \frac{2\pi}{\gamma_{\mu}} \nu_{\mu}$) $\approx$ 0.1\,mT produced by the adjacent Ru$^{3+}$ spins.~\cite{PhysRevB.95.121111} This makes $\mu$SR an ideal probe to detect the presence of any tiny static magnetism. It is clear that the present ZF-$\mu$SR spectra display the characteristic signals originating from static magnetism: 1.) Spontaneous coherent oscillations in the studied temperature range down to 1.5\,K.,  
2.) Strong damping of the muon depolarization due to the random distribution of the static field. These points demonstrate the presence of a well-defined static magnetic field at the muon stopping site originating due to the possible long-range magnetic ordered state of Ru$^{3+}$ moments. 
\\
\indent
The ZF-time spectra can be best physically described by two magnetic and one nonmagnetic component:

\begin{align}
A(t)= \nonumber A_1(\frac{2}{3}\cos(\omega_{H1}t+\phi)e^{-\lambda_\mathrm{T1}t}+\frac{1}{3}e^{-\lambda_\mathrm{L1}t}) \\ \nonumber
+ A_2(\frac{2}{3}\cos(\omega_{H2}t+\phi)e^{-\lambda_\mathrm{T2}t}+\frac{1}{3}e^{-\lambda_\mathrm{L2}t}) \\
+A_3e^{-\lambda_\mathrm{L2}t}
 \label{eqn:muonAsymmetry}
\end{align}
where $A_{i}$(i=1,2,3) represent the initial asymmetry. 
$\lambda_{Ti} (i=1,2)$, $\lambda_{Li} (i=1,2)$ 
are the width of the static field distribution and muon relaxation rate, respectively.
$\omega_{H2}$ can be well described by the following equation: $\omega_{H2}(T)$ = a $\omega_{H1}(T)$, where a is a temperature independent factor. Thus both muon sites observe the same physical behavior, unlike the observations in $\alpha$-RuCl$_3$.~\cite{PhysRevB.94.020407-rucl3}  
   
\begin{figure}[h]
\includegraphics[width=\columnwidth]{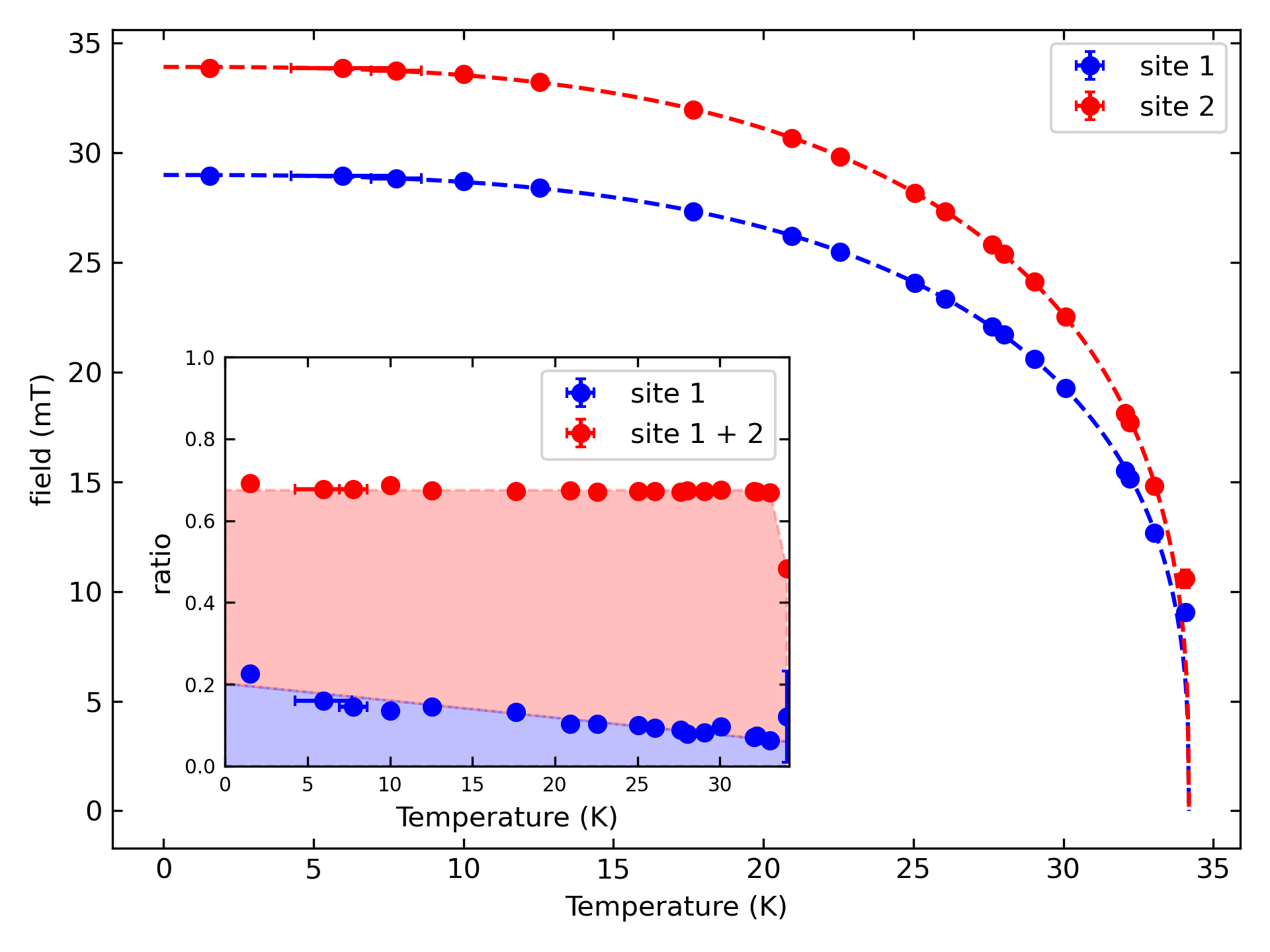}
\caption{\label{fig:order-parameter} Temperature dependence of the internal fields by fitting the ZF spectra. Dotted lines are ordered parameter fit as detailed in the main text. Inset shows the amplitude ratios of the ordering fractions estimated from the fitting of the time spectra.}
\end{figure}

The right panel of Fig.~\ref{fig:ZF-musr-spectra} shows the temperature dependence of Fourier-transformed $\mu$SR time spectra in zero fields. In the ordered state field spectra just below $\approx$ 34\,K shows a single broad line. With lowering the temperature lines are shifted toward the high field side and clearly show the split field spectra suggesting two well-defined internal fields at the muon sites indicated by the dashed lines.
\\
The estimated internal fields from the fit of the time spectra as a function of temperature are plotted in Fig.~\ref{fig:order-parameter}. This order parameter plot for site 1 and site 2 shows that at $\approx$ 34\,K both fields merge indicating one magnetic ordering temperature of the system. However, lowering the temperature they split, and further lowering the temperature they are saturated to different fixed values. The overall behavior of the order parameter development appears to be similar suggesting similar magnetism of the two sites.  The obtained parameters can be described by the following phenomenological equation:
\begin{align}
B(T)= B_0 {[1-(\frac{T}{T_N})^\alpha ]}^\beta
\label{eqn:muonAsymmetry}
\end{align} 
The fits give the following parameter values of 
$B_0$($\mu_0H_0$)=28.99(2), $\alpha$=2.81(4), $\beta$=0.35(2) and $T_N = 34.19(7)$\,K. These fits identified the magnetic ordering temperature to be $T_N = 34.19(7)$\,K which is consistent with the bulk characterization, neutron scattering, and NMR experiments. Let's briefly discuss the meaning of all the parameters used in the phenomenological order parameter equation and correspondingly relevant physics that one can extract from there. $B_0$ is a constant that determines the saturated magnetic field. $\beta$ is called the critical exponent and gives information about the type of magnetic ordering. For three-dimensional Heisenberg ordering the expected value of $\beta$ is 0.367, whereas for three-dimensional Ising magnets the expected $\beta$ is 0.326.~\cite{{Kuo-Feng-2016},{PhysicsReports2002}}  The current experimental finding, in particular $\beta$, indicates that the magnetic ordering in RuBr$_3$ is more like a quasi-three-dimensional type.

For analog material $\alpha$-RuCl$_3$ the four potential muon stopping sites have been estimated.~\cite{PhysRevB.94.020407-rucl3} Given that RuBr$_3$ has a similar structure to $\alpha$-RuCl$_3$, it is likely the similar muon-stopping sites for RuBr$_3$. Both in $\alpha$-RuCl$_3$ and RuBr$_3$ two distinct well-resolved peaks are observed in the field spectrum in the ordered state. While deep in the ordered state, the development of the ordered parameter qualitatively behaves similarly for both compounds, the significant differences are noticeable at the onset of the ordering temperature.
Further $\mu$SR studies on $\alpha$-RuCl$_3$ single crystals show multiple magnetic phases due to the presence of stacking faults.~\cite{PhysRevB97134410}
 In the case of RuBr$_3$ magnetic ordering sets in at the same temperature, whereas in $\alpha$-RuCl$_3$ the ordering sets in at two different temperature points. One likely scenario is that in the former case, the presence of less amount of stacking faults makes a unique ordering temperature for the potential muon-stopping sites. The anisotropy in the interaction may play a significant role there. However, to what extent this is related to the anisotropy of the material is not clear so far. This would be clarified by investigating single crystals which are currently not available to our knowledge.

In the inset of Fig.~\ref{fig:order-parameter}, the amplitude ratios of the ordering fractions are displayed as estimated from the fitting of the time spectra. The two signal fractions are not equal, but rather significantly different. In the solid-state Nuclear Magnetic Resonance (NMR) technique to evaluate field/frequency sweep spectra, the area under the curve is taken as the measure of the number of nuclei effectively contributed to the signal. Thus, the signal intensity and/or area under the curve is proportional to the number of nuclei. This helps to quantify the relevant signal fraction from different surroundings of the nuclei. Conceptually, the muon field spectra are not that different. Significantly intensity differences suggest among the four potential muon-stopping sites one-third of the muons (site 2) has its different environments than the rest (site 1).  Although most of the muons encounter qualitatively similar magnetic field distribution because of the more three-dimensional magnetic ordering of the Ru$^{3+}$ spins, their different locations in the lattice with respect to the Ru$^{3+}$ spins simply encounter different magnetic field values.

\begin{figure}[h]
\includegraphics[width=\columnwidth]{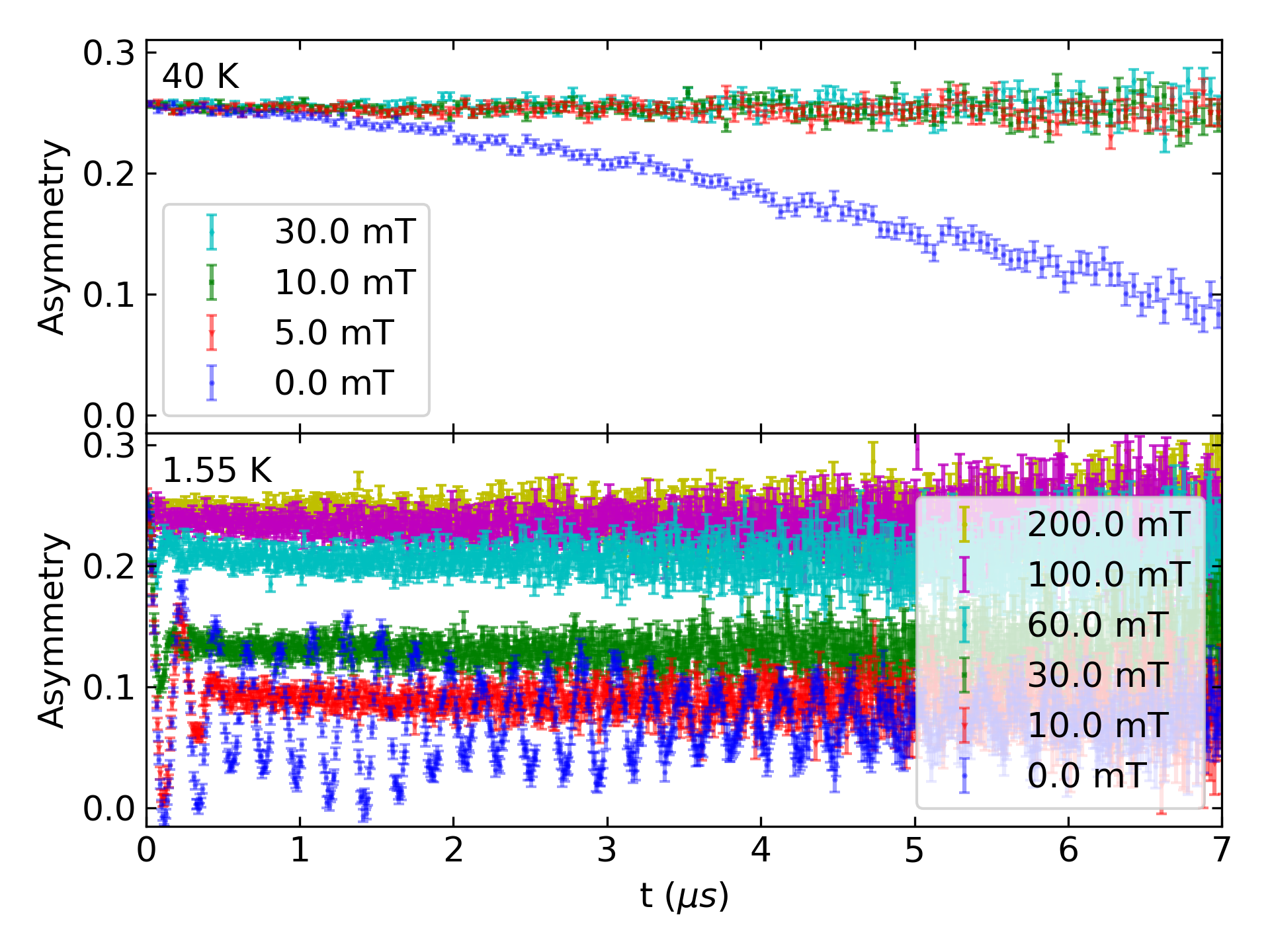}
\caption{\label{fig:LF-spectra} Top panel shows the representative LF-\muSR\  time spectra collected at 40\,K. The lower panel shows the LF spectra at the selected field at 1.55\,K.}
\end{figure}

Having established that the system has the three-dimensional magnetic ordering of the Ru$^{3+}$ spins, we now want to understand the character of the fluctuations of the Ru$^{3+}$ spins, namely whether the fluctuations of the spins are static or dynamic in nature. In general, for a QSL ground state system, the spins are fluctuating, even if there is static ordering, and this is reflected in the
\muSR\ relaxation processes by means of persistent spin dynamics or constant values of relaxation in low temperature. LF-\muSR\ experiments are an excellent test case to prove or disprove the character of the spin fluctuation (static/dynamic). Figure~\ref{fig:LF-spectra} shows the LF-$\mu$SR time spectra at two different temperatures, namely 40\,K and 1.55\,K representing the normal state and magnetically ordered state, respectively. In the normal state at 40\,K by applying a small field of only 50 Gauss, it is possible to completely decouple the time spectra. The relaxation process at zero field appears to be the effect of static nuclear relaxation which is easy to suppress/decouple by a small amount of longitudinal field. So no electronic contribution either static or dynamic has been encountered by the muons. However, deep in the ordered state (lower panel) time spectra can be decoupled by the application of the longitudinal fields reflecting the absence of dynamics.

\begin{figure}[h]
\includegraphics[width=\columnwidth]{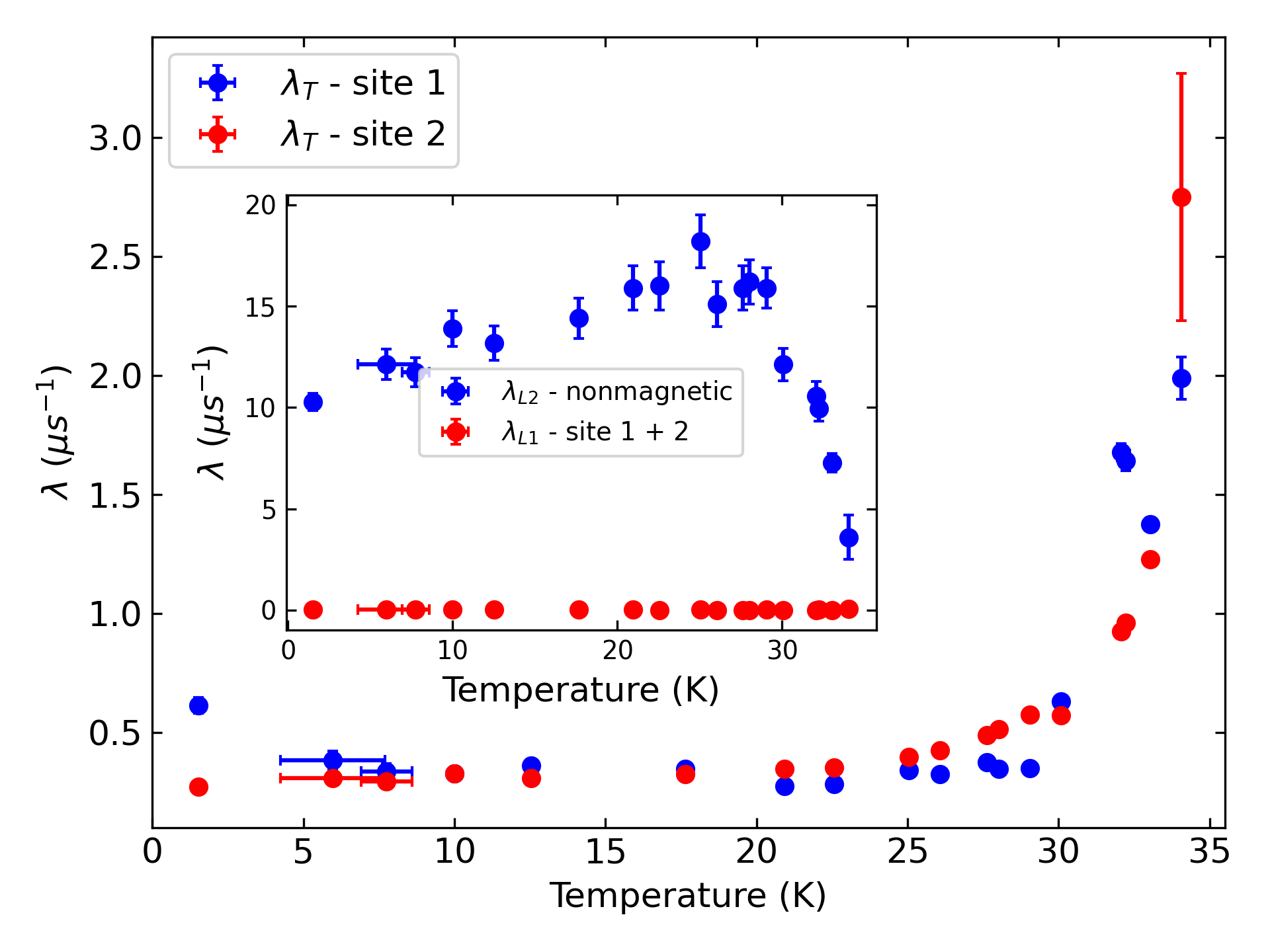}
\caption{\label{fig:relaxation-rate} Temperature dependence of the ZF transverse relaxation rates in the ordered state. The inset shows the ZF longitudinal relaxation rates.}
\end{figure}

Figure~\ref{fig:relaxation-rate} shows the ZF transverse ($\lambda_T$) and longitudinal relaxation ($\lambda_L$) (inset) rates in the magnetically ordered state. Both relaxation processes clearly represent a static ground state. In the ordered state for the magnetic component, $\lambda_L$ shows the nominal temperature dependency with values close to zero. Both relaxation processes clearly represent a static ground state.~\cite{PDalmasdeReotier1997}

\textit{Conclusions.} 
In conclusion, detailed \muSR\ studies on the powder of RuBr$_3$ are carried out. There is clear evidence of long-range magnetic ordering below $\approx$ 34\,K. \muSR\ relaxation rate $\lambda$ values below the similar temperature range are typical for a static magnetically ordered system. This is in line with the decoupling experiments. Taken all together, RuBr$_3$ shows three-dimensional long-range magnetically ordering at least in zero fields in \muSR\ time window.  However, unlike the case of $\alpha$-RuCl$_3$ in the \muSR\ experimental window, where full three-dimensional order sets in below a certain temperature after the first ordering takes place, in RuBr$_3$ three-dimensional order sets in just at $\approx$ 34\,K. The problem remains open to what would be the anisotropic magnetic properties in this system in external magnetic fields, for which one needs single crystals.

\textit{Acknowledgments.} This work was financially supported by the Deutsche Forschungsgemeinschaft (DFG) within the SFB 1143 “Correlated Magnetism – From Frustration to Topology”, project-id 247310070 (Projects C02), GR\,4667/1. This work was supported by JST, CREST (JPMJCR1901), and JSPS KAKENHI Nos. JP22H00102, JP19H05823, and JP19H05822. We thank Stephen Blundell for the helpful discussions.

\bibliography{RuBr3}

\end{document}